\documentclass[oldversion,letter]{aa}
\usepackage{txfonts}
\usepackage{graphicx}
\usepackage[]{natbib}
\bibpunct{(}{)}{;}{a}{}{,}

\begin{document}

\title{The nature of the fluorescent iron line in V~1486~Ori}
\author{S. Czesla \and J.H.H.M. Schmitt}
\institute{Hamburger Sternwarte, Universit\"at Hamburg, Gojenbergsweg 112, 21029 Hamburg, Germany}
\date{Received ... / accepted ...}
\abstract
{
  The fluorescent $6.4$~keV iron line provides information on cool material in the
  vicinity of hard X-ray sources as well as on the characteristics of the X-ray sources themselves. First discovered in
  the X-ray spectra of the flaring Sun, X-ray binaries and active galactic nuclei (AGN), the fluorescent line was also
  observed in a number of stellar X-ray sources.
  The young stellar object (YSO) V1486~Ori was observed in the framework of the \textit{Chandra} Ultra Deep Project (COUP) as the
source COUP~331. We investigate its spectrum,
  with emphasis on the
  strength and time variability of the fluorescent iron K$_{\alpha}$ line,
  derive and analyze the light curve of COUP~331 and proceed with a time-resolved spectral analysis of the
  observation.
  The light curve of V~1486~Ori shows two major flares, the first one lasting for $\approx 20$~ks with a peak X-ray luminosity of 
  $2.6\cdot 10^{32}$~erg/s (dereddened in the $1-10$~keV band) and the second one -- only partially observed -- for $>60$~ks with 
  an average X-ray luminosity of $2.4\cdot 10^{31}$~erg/s (dereddened).
  The spectrum
  of the first flare is very well described by an absorbed thermal model at high temperature, with a pronounced 6.7~keV
  iron line complex, but without any fluorescent K$_{\alpha}$ line.  The X-ray spectrum of
  the second flare is characterized by even higher temperatures ($\gtrsim 10$~keV)
  without any detectable 
  $6.7$~keV Fe~XXV feature, but with a very strong fluorescent iron K$_{\alpha}$ line appearing predominantly in the 20~ks
  rise phase of the flare. Preliminary model calculations indicate that photoionization is unlikely to
  account for the entire fluorescent emission during the rise phase.
}
\keywords{stars: early-type, activity -- X-ray: stars }
\maketitle

\section{Introduction}
\vspace{-0.6cm}
 The COUP data set provides a nearly continuous 13~day long X-ray observation of the Orion nebula
 star forming region and is currently one of the best suited data sets for the exploration of the X-ray properties of 
 large and homogeneous samples of young stars.  Substantial analysis of this data set has already been performed
 \citep[e.g. by][]{getman_2005}. In particular, \citet{tsujimoto_2005} carried out a systematic search for the occurrence of the
 fluorescent iron K$_{\alpha}$ line at 6.4~keV,
 identifying $7$~COUP sources with detectable fluorescent emission. 
 This emission line is particularly interesting since it can provide a wealth of information on the emitting source(s). 
 Fluorescent $6.4$~keV photons -- hereafter referred to as Fe~I K$_{\alpha}$ photons -- result from a de-excitation of neutral or lowly ionized 
 iron atoms, following an excitation leaving the iron ion with a K-shell vacancy.
 When the latter is filled by another (mostly L-shell) electron, the excess energy can be 
 carried by a fluorescent photon. Highly ionized iron can also show this behavior, but the transition energy
 rises because of the reduced screening of the inner electrons from the nuclear charge.
 Several processes lead to the ejection of a K-shell electron and, thus, to the creation of the 
 Fe~I K$_{\alpha}$ line.  During the process of photoionization a K-shell electron is ejected from iron after the
 absorption of a photon with an energy exceeding $7.11$~keV.  Alternatively, high energy particles can
 eject K-shell electrons with the same threshold energy.
 
 \noindent
 The first detection of the fluorescent Fe~I K$_{\alpha}$ line was reported in the context of solar physics
 \citep[e.g.][]{neupert_1967}. A model succeeding in explaining most of the solar Fe~I K$_{\alpha}$ emission 
 with photoionization only was proposed and evaluated by \citet{bai_1979}, yet, 
 e.g., \citet{zarro_1992} argue that photoionization alone is not sufficient to account for  
 all of the solar Fe~I K$_{\alpha}$ emission. Fluorescent K$_{\alpha}$ line emission has also been reported from other celestial
 X-ray sources. It is typically encountered in the high-energy emission from AGN and X-ray binaries,
 which are characterized by high X-ray luminosities and rather hard X-ray spectra, thus providing large photon numbers 
 above the $7.11$~keV threshold for photoionization. \citet{fabian_1989} detect the fluorescent iron line in the spectrum
 of the X-ray binary Cyg-X1, and \citet{tanaka_1995} were the first to report the relativistically broadened
 line in the spectrum of the AGN MCG-6-30-15.  Recent observations with XMM-{\it Newton} and {\it Chandra} extended the
 class of Fe~I K$_{\alpha}$ line emitters to a number of stellar sources, 
 such as the class~I YSO
 Elias~2-29 
 \citep{favata_2005}, the above seven COUP sources discussed
 by \citet{tsujimoto_2005}, and the YSO YLW~16A for which \citet{imanishi_2001} carried out 
 time-resolved spectroscopy. 
 
 In this letter we present a detailed analysis of the X-ray properties of the YSO V~1486~Ori (= COUP~331) with emphasis on
 its Fe~I K$_{\alpha}$ emission, which was already reported by \citet{tsujimoto_2005}, who measured
 a normalization of $2.6(0.1-4.6)\cdot 10^{-7}$~ph/(cm$^2$ s) with an equivalent width of $126$~eV for the fluorescent line.
 Unfortunately, the YSO  V~1486~Ori has so far not received much attention at 
 other energy bands. From the 2MASS-all-sky survey we infer its infrared magnitudes as $12.5$~mag, $10.6$~mag,
 and $9.4$~mag, in the J,H, and K$_s$ bands, respectively. \citet{tsujimoto_2005} show that the infrared
 colors of V~1486~Ori do not match the regions of reddened dwarfs or giants and argue for the presence
 of a NIR-emitting inner disk.  Moreover a rotation period of $(6.09\pm 0.3)$~d is known for V~1486~Ori (\citet{herbst_2002}
 and \citet{stassun_1999}). Following \citet{tsujimoto_2005} we 
 adopt a distance of $450$~pc for Orion.

  \vspace{-0.35cm}
\section{Observations and data analysis}
 \label{sec:ObsData}
 \vspace{-0.25cm}
 The COUP data was obtained during 13 consecutive days in Jan.~2003 with the Advanced CCD Imaging Spectrometer (ACIS) on board the \textit{Chandra}
 X-ray observatory. The total exposure
 time is \mbox{$\approx 840$~ks}, separated into 6 observation segments extending over a
 total time of $\approx 1140$~ks.  All observations have the same aimpoint
 ($\alpha=5^h35^m17^s$ and $\delta=-5^{\circ}23'40''$) 
 and the field of view covers an area of $17'\times 17'$;  the source COUP~331 is located at
 $\alpha =5^h35^m9.2^s$ and $\delta=-5^{\circ}30'58''$ \citep[cf.][]{getman_2005}.   
 Our analysis is based on the pre-processed photon
 data provided by the \textit{Chandra} pipeline; for further analysis we applied the 
 CIAO-software in version~3.4.  We first screened the photon data for events in the $0.3-13$~keV energy
 band to suppress background. Thereafter, we defined source and background regions as a circle with an $8$~arcsec radius
 and an annulus extending from $\approx 14$-$32$~arcsec, both centered on the nominal source position.
 \vspace{-0.25cm}

 \subsection{The light curve}
\begin{figure}[h]
    \centering
    \includegraphics[width=0.49\textwidth]{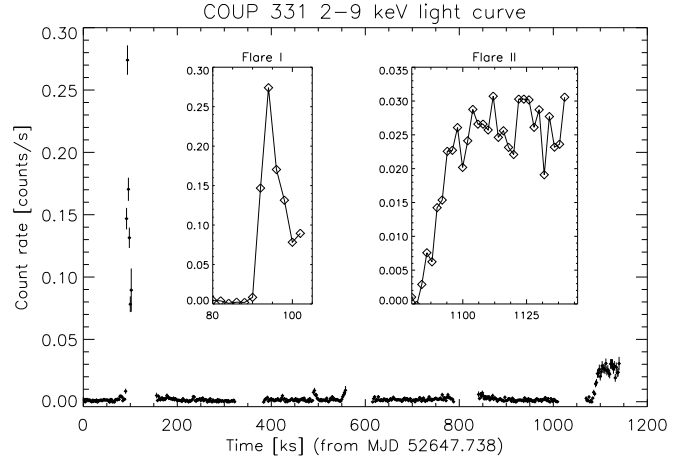}
    \caption{Background-subtracted light curve of COUP~331 with 2~ks binning. 
    Inserted panels show close-up views of the flares I and II. 
    \label{fig:LightCurve}}
 \vspace{-0.5cm}
  \end{figure}

  \vspace{-0.25cm}
  We generated the background-subtracted $2-9$~keV band light curve of COUP~331 shown 
  in Fig.~\ref{fig:LightCurve}. The 6~observation segments can be clearly
  identified. For most of the COUP observations COUP~331 was recorded at a count rate of $3.3\cdot 10^{-3}$~cts/s,
  except for two periods.  A strong flare (termed ``flare~I") occurred $\approx 90$~ks after 
  the beginning of the observation. The flare rapidly rises to its peak within $\approx 4$~ks, the decay
  is again quite rapid with an e-folding time of $\tau_{decay}=5.2$~ks, but, unfortunately, is only partially observed.
  A second, less prominent but very significant count rate enhancement
  (termed ''flare II") occurred 1080~ks after the start of the COUP campaign. This second flare
  was also not fully covered.  Flare~II has a much slower rise phase lasting for $\approx 20$~ks, and
  thereafter, the $2-9$~keV band count rate remains more or less constant for at least $40$~ks when the COUP
  observations were terminated.  For the rest of the observations COUP~331 was found in quiescence.  
  \vspace{-0.3cm}
  
  \subsection{Spectral analysis}
   \vspace{-0.25cm}
   In the following section we carry out temporally resolved spectral analysis of the
   flare~I, flare~II, and quiescent phases of COUP~331; note that all quoted errors
   refer to $90$\% confidence intervals unless stated otherwise.
   COUP~331 is located about $7$~arcmin off-axis, hence it does not appear as a perfectly
   symmetric point-source. However, we regard its image distortion as weak and apply the CIAO standard tools
   for the extraction of point-source spectra (i.e.~\texttt{psextract}).
   The spectra were analyzed using the XSPEC environment in version 11.3.1.

  \subsubsection{The quiescent phase}
    \vspace{-0.25cm}
    \label{sec:QuietSpec}
    The quiescent phase spans the $0-1050$~ks time interval with the exception of the first flare event, and is
    the longest lasting phase, covering $\approx 90$\% of the total observation time. Note that we disregard
    a $\approx 20$~ks contribution preceding flare~II. During quiescence COUP~331 is detected with a net count-rate of 
    $3.3\cdot 10^{-3}$~cts/s corresponding to an (unabsorbed) X-ray luminosity of $1.6\cdot 10^{30}$~erg/s.
    We checked for temporal
    changes in the quiescent spectrum, but did not find evidence for any significant variations.
    The quiescent X-ray spectrum is well described by an absorbed thin-thermal plasma 
    model (we used the VAPEC models) with all metal abundances -- apart from Fe, which is
    left as a free parameter -- fixed to $0.3$ times the solar values. In Fig.~\ref{fig:FlareISpec} we show the data and the fit;
    the best fit model parameters are presented in Table~\ref{tab:FitPara1}, where we list the
    derived absorbing hydrogen column-density, the temperature,
    the iron abundance and the fit quality.
    \vspace{-2mm}
  
  \subsection{Flare~I}
\begin{figure}[h]
      \centering
      \includegraphics[width=0.49\textwidth]{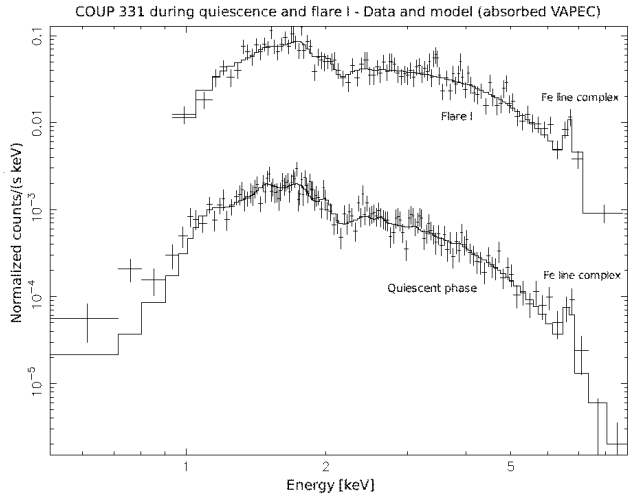}
      \caption{Spectrum of COUP~331 during flare~I (upper curve)  and the quiescent phase (lower curve), both fitted 
      with an absorbed VAPEC model.
      \label{fig:FlareISpec}}
 \vspace{-0.3cm}
    \end{figure}

    \vspace{-0.3cm}
    For an analysis of flare~I we considered all data recorded in the time interval from $87.5$ to $102$~ks; the
    spectrum of flare~I is shown in Fig.~\ref{fig:FlareISpec}.
    The flare spectrum is very well described by the same model as applied for the quiescent phase
    (cf.~Sect.~\ref{sec:QuietSpec}) with somewhat different spectral parameters listed in Table~\ref{tab:FitPara1}.
    We find -- as expected for a flare -- an enhanced temperature of $6.8$~keV, leading to a spectrum with a clearly
    detected Fe line feature at 6.7 keV, but with no detected excess emission at $6.4$~keV.
    From the spectrum we derive
    an (unabsorbed) peak X-ray luminosity of L$_\mathrm{X}=2.6\cdot 10^{32}$~erg/s (in the $1-10$~keV band) reached directly after the rise phase,
    and -- with an e-folding time of $5.2$~ks for the decay phase -- 
    a total energy output of $\Delta E\approx 1.4\cdot 10^{36}$~erg in the same band.  With these numbers, flare I observed 
    on COUP~331 is among the largest stellar flares ever observed.
    
  \subsection{Flare~II}
\begin{figure}[h]
      \centering
      \includegraphics[width=0.49\textwidth]{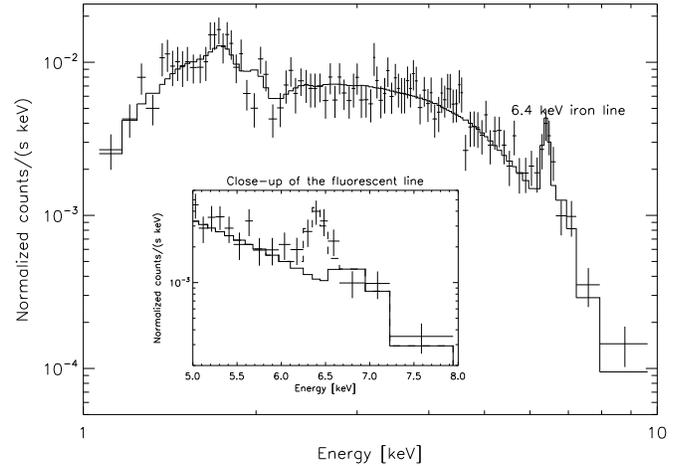}
      \caption{The COUP~331 spectrum during flare~II with a close-up of the $5-8$~keV energy band.
      Two model fits are indicated. First, (solid
      line) an absorbed thermal model and second (dashed line), the same thermal model with additional 
      instrumentally-broadened Gaussian component centered at $6.4$~keV.
      \label{fig:FluoSpec}}
    \vspace{-0.2cm}
    \end{figure}

    \vspace{-0.25cm}
    For the analysis of flare~II we considered all data recorded after $1080$~ks (cf. Fig.~\ref{fig:LightCurve}).
    The X-ray spectrum was generated and
    analyzed in the same fashion as those of the quiescent and flare~I phases.  
    From the spectrum of
    the entire observed flare~II phase (see Fig.~\ref{fig:FluoSpec}) we calculate an $1-10$~keV band X-ray
    luminosity of $2.4\cdot 10^{31}$~erg/s. Considering a duration of at least $60$~ks, the total energy output in
    this energy band is $\Delta E > 1.4\cdot 10^{36}$~erg.
    
    In the fit process of
    the flare~II data we encountered two problems. 
    First, the absence of temperature-sensitive features in the spectrum -- especially the hot Fe~XXV feature at $6.7$~keV 
    and its Fe~XXVI counterpart near $7$~keV -- 
    prevents us from tightly constraining the plasma temperature. We find
    that any choice of the temperature beyond $10$~keV leads to acceptable fits, and thus decided to apply a value
    of $13$~keV.
    Second, the spectrum shows definite excess 
    emission above the thermal continuum at $6.4$~keV, which we attribute to the Fe~I K$_{\alpha}$ line.
    We therefore included an additional instrumentally broadened Gaussian component centered at 6.4~keV into our model
    and refitted, leaving only the normalization of the Gaussian component free to vary. The thus obtained fit is shown
    in Fig.~\ref{fig:FluoSpec}, which shows the obvious presence of the 6.4~keV line and 
    virtual absence of the $6.7$~keV Fe~XXV feature.
    The resulting fit parameters for the thermal model are again listed in 
    Table~\ref{tab:FitPara1}, and for the Gaussian component we arrive at a normalization of 
    \mbox{$5(3.3-6.7)\cdot 10^{-6}$~ph/(cm$^2$ s)}, corresponding to an equivalent width of \mbox{685(452-918)~eV}. As is obvious from Fig.~\ref{fig:FluoSpec}, the inclusion of
    a Gaussian component at $6.4$~keV into the model significantly 
    increases the fit quality. 
    \vspace{-0.5cm}

    \begin{table}[h!]
    \begin{minipage}[h]{0.5\textwidth}
    \renewcommand{\footnoterule}{}
      \caption{Fit parameters for the absorbed VAPEC model.
      \label{tab:FitPara1}}
      \begin{tabular}{l | l l l l}
        Phase        & n$_{\mathrm{H}}$             & Temp.                &   Ab$_{\mathrm{Fe}}$        & $\chi^2$/d.o.f.   \\
                     & $[10^{22}\mathrm{~H/cm}^2]$  & [keV]                &   $[\mathrm{Ab}_{\sun}]$    &                   \\ \hline \hline
        Quiescent    & $1.09_{0.97}^{1.21}$         & $3.0_{2.6}^{3.4}$    & $0.32_{0.12}^{0.54}$        & 131.6/134         \\[0.5mm]
        Flare~I      & $1.28_{1.18}^{1.41}$         & $6.8_{5.5}^{8.2}$    & $0.73_{0.46}^{1.02}$        & 94.2/99           \\[0.5mm]
        Flare~II     & $1.55_{1.41}^{1.66}$         & $\gtrsim 13$         & $\approx 0.88$\footnote{This value is ill-constrained since temperature-dependent}         & 102/95            \\
      \end{tabular}
      \vspace{-3mm}
    \end{minipage}
    \end{table}
    \vspace{-0.4mm}

\begin{figure}[h]
      \centering
      \includegraphics[width=0.49\textwidth]{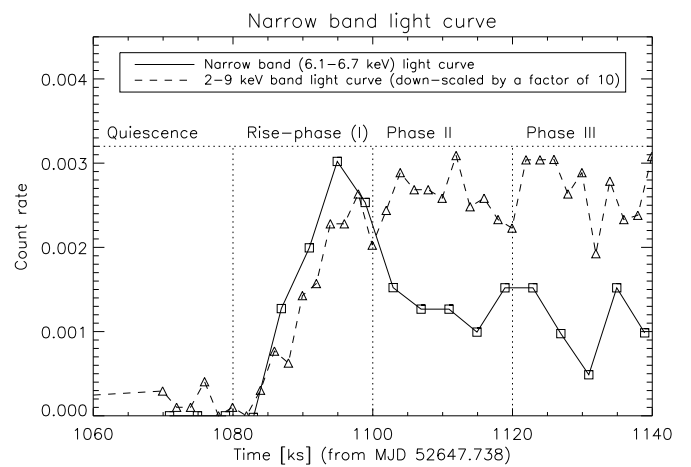}
      \caption{The background-subtracted narrow-band light curve (solid line) of COUP~331 during the last 80~ks of the COUP observation. The
      light curve peaks in the rise-phase of flare~II. For
      clarity the $2-9$~keV light curve is also shown (dotted line - scaled to fit the plot range).
      \label{fig:NarrowBandLC}}
    \end{figure}

\begin{figure}[h]
      \centering
      \includegraphics[width=0.49\textwidth]{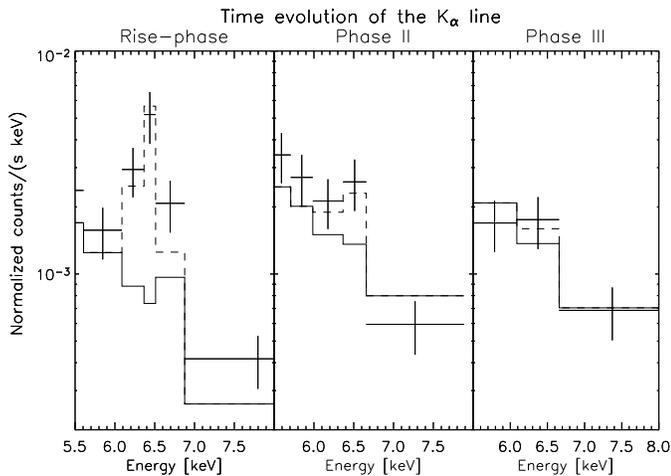}
      \caption{Time-resolved $5.5-8$~keV section of the COUP~331 spectrum
      (cf. Fig.~\ref{fig:NarrowBandLC}). Note the decrease of the Fe~I K$_{\alpha}$ line's strength.
      \label{fig:FlareII3Phase}}
 \vspace{-1.0cm}
    \end{figure}

    \vspace{-0.5cm}
    \noindent
    As a first result we thus note that the 6.4~keV emission feature is present -- at best -- only during 60~ks during
    flare~II.  In order to further temporally constrain the presence of the 6.4~keV iron line we constructed a background-subtracted
    $6.1-6.7$~keV narrow-band light curve of COUP~331 with a $4$~ks binning in addition to the $2$~ks binned
    $2-9$~keV band light curve; both light curves are shown in Fig.~\ref{fig:NarrowBandLC}.  As is clear from 
    Fig.~\ref{fig:NarrowBandLC} the $2-9$~keV band light curve stays more or less constant after the rise to peak, while
    the $6.1-6.7$~keV narrow-band light curve decays quite rapidly. We therefore subdivided the 
    the flare~II phase data into three consecutive $20$~ks time intervals, denoted by ``rise-phase", ``phase~II", and ``phase~III"
    as indicated in Fig.~\ref{fig:NarrowBandLC}.
    \vspace{-0.5cm}
    The X-ray spectra in the 5-8 keV region corresponding to these time intervals are shown 
    in Fig.~\ref{fig:FlareII3Phase}. In these plots
    the solid line indicates the thermal model described earlier. For all spectra the same model was applied, but we allowed
    the normalizations of the VAPEC and Gaussian component to be fitted independently for all time intervals; 
    all other parameters were regarded globally. Figure~\ref{fig:FlareII3Phase} clearly demonstrates an evolution of
    the Fe~I K$_{\alpha}$ line strength. Interestingly the line normalization decreases as time proceeds, while the
    underlying thermal spectrum remains almost unaffected. We thus interpret the variations in the narrow-band light curve
    shown in Fig.~\ref{fig:NarrowBandLC} as arising from variations in the Fe~I K$_{\alpha}$ feature's strength. Note that
    the constancy of the thermal spectrum does not imply the same for the temperature, since the latter exceeds
    $10$~keV, producing a flat spectrum in the ``low energy" band we are observing.
    The K$_{\alpha}$ feature is strongest during the flare~II rise phase where our fits yield a line flux of
    $9.1(5.8-13.4)$~$10^{-6}$~ph/(cm$^2$ s) (an EW of $\approx 1400$~eV) while
    in phase II and III it is reduced
    to 3.04(0.8-7.6) and 0.9(0-4.8)$ \cdot 10^{-6}$~ph/(cm$^2$ s), respectively.
    This is also reflected by
    the narrow-band light curve around $6.4$~keV (cf. Fig.~\ref{fig:NarrowBandLC}). 
    \vspace{-2mm}

  \section{Interpretation and conclusions}
   \label{sec:Results}
   \vspace{-0.25cm}
   Our time resolved analysis of the COUP X-ray data of the YSO COUP~331 shows that for
   $\approx 90$\% of the entire observation time the object is found in a quiescent state characterized by
   an X-ray temperature of $30-40$~MK.    Two major flares are covered by the 
   COUP observation. In the first data segment
   a relatively short duration event with a peak X-ray luminosity of $2.6\cdot 10^{32}$~erg/s
   occurred, which clearly shows the $6.7$~keV iron feature, but {\bf no}
   Fe~I K$_{\alpha}$ emission.

   In the last COUP data segment a long duration flare event occurred; starting with a
   $\approx 20$~ks rise phase it lasted $>60$~ks. The decay of this flare cannot be constrained because the 
   COUP observations were stopped.   The total amount of energy released in the $1-10$~keV energy band 
   is $> 1.4\cdot 10^{36}$~erg;
   the X-ray temperature of this flare is extremely large ($\gtrsim 10$~keV)
   and can in fact not be well constrained with the {\it Chandra} data. 
   A clear and highly significant detection of the fluorescent K$_{\alpha}$ line is obtained only during the $20$~ks rise
   time of this flare; at later times the line may still be present albeit at a lower level.
   This event on COUP~331 resembles a huge flare observed on 
   Algol, observed by \citet{favata_1999} with BeppoSAX, which also shows a slow rise phase, followed by
   a nearly constant phase - lasting for $\approx 80$~ks - in its $1.6-10$~keV band light curve.  However, 
   the corresponding $15-100$~keV band light curve clearly shows the characteristics of a decay phase 
   (cf., \citet{favata_1999}, Figs.1 and 2). The energy output of $1.4\cdot 10^{37}$~erg derived by 
   \citet{favata_1999} in the $0.1-10$~keV band compares well to the numbers estimated for COUP~331, note however
   that no Fe~I K$_{\alpha}$ line was detected.
   
   The precise temporal association between Fe~I K$_{\alpha}$ line emission and high energy continuum is necessary in
   order to quantitatively assess the origin of 
   the Fe~I K$_{\alpha}$ emission observed in COUP~331; we find that the actual photon flux
   in the fluorescent K$_{\alpha}$ line during the flare~II rise phase is $35$~times larger than reported by  
   \citet{tsujimoto_2005}. This discrepancy arises because \citet{tsujimoto_2005} give time-averaged values,
   whereas our values refer to several limited time intervals.
   The X-ray temperature during
   the flare event was extremely large; the total luminosity in the Fe~I K$_{\alpha}$ line alone is $\approx 2.3\cdot 10^{30}$~erg/sec.
   While these circumstances appear to favor the excitation of the Fe~I K$_{\alpha}$ line 
   through photoionization, preliminary model calculations challenge such a scenario.
   We repeated the calculations carried out by \citet{bai_1979} applying a reflector with cosmic abundances. We used the same branching ratios
   and cross sections for iron, but slightly different cross section for absorption by other elements composed according to 
   cosmic abundances.
   Furthermore, we located the source directly on the reflector to make it subtend half the sky, thus providing
   an upper limit for the fluorescent iron line flux. The illuminating input spectrum was modeled as a 
   thermal X-ray spectrum with a temperature of $13$~keV, normalized to the measured fluxes. 
   Our preliminary calculations show that an already optimally assumed fluorescence geometry significantly underpredicts the observed
   Fe~I K$_{\alpha}$ flux, making an  
   interpretation of the Fe~I K$_{\alpha}$ line through photoionization difficult to accept.
   However, these calculations will be extensively discussed in a larger and systematic context in a forthcoming paper, 
   where we will also address its impact
   on the characteristics of COUP~331 and other sources with fluorescent K$_{\alpha}$ emission. 
   \vspace{-1mm}
   \begin{acknowledgements}
     SC acknowledges support from the DLR under grant 50OR0105.
   \end{acknowledgements}
   
  \bibliographystyle{aa}
   \vspace{-0.59cm}
  \bibliography{7741}

\end{document}